\documentclass[fleqn]{llncs}
\usepackage{sou}

\title{On the Operating Unit Size\\ of Load/Store Architectures%
\thanks{This research was partly carried out in the framework of the
        GLANCE-project\protect\linebreak[3] MICROGRIDS, which is funded
        by the Netherlands Organisation for Scientific Research (NWO).}}

\author{J.A. Bergstra \and C.A. Middelburg}
\institute{Programming Research Group,
           University of Amsterdam, \\
           P.O.~Box~41882, 1009~DB~Amsterdam, the Netherlands \\
           \email{J.A.Bergstra@uva.nl, C.A.Middelburg@uva.nl}
           }

\begin{document}
\maketitle

\begin{abstract}
We introduce a strict version of the concept of a load/store instruction
set architecture in the setting of Maurer machines.
We take the view that transformations on the states of a Maurer machine
are achieved by applying threads as considered in thread algebra to the
Maurer machine.
We study how the transformations on the states of the main memory of a
strict load/store instruction set architecture that can be achieved by
applying threads depend on the operating unit size, the cardinality of
the instruction set, and the maximal number of states of the threads.
\end{abstract}

\section{Introduction}
\label{sect-introduction}

In~\cite{BM05f}, we introduced Maurer machines, which are based on the
model for computers proposed in~\cite{Mau66a}, and extended
basic thread algebra, which is introduced in~\cite{BL02a} under the name
basic polarized process algebra, with operators for applying threads to
Maurer machines.
Threads can be looked upon as the behaviours of deterministic sequential
programs as run on a machine.
By applying threads to a Maurer machine, transformations on the states
of the Maurer machine are achieved.
In~\cite{BM06b}, we proposed a strict version of the concept of a
load/store instruction set architecture for theoretical work relevant to
the design of micro-architectures (architectures of micro-processors).
We described the concept in the setting of Maurer machines.
The idea underlying it is that there is a main memory whose elements
contain data, an operating unit with a small internal memory by which
data can be manipulated, and an interface between the main memory and
the operating unit for data transfer between them.
The bit size of the operating unit memory of a load/store instruction
set architecture is called its operating unit size.

In this paper, we study how the transformations on the states of the
main memory of a strict load/store instruction set architecture that can
be achieved by applying threads to it depend on the operating unit size,
the cardinality of the instruction set, and the maximal number of states
of the threads.
The motivation for this work is our presumption that load/store
instruction set architectures impose restrictions on the expressiveness
of computers.
Evidence of this presumption is produced in the paper.
In order to present certain results in a conveniently arranged way, we
introduce the concept of a thread powered function class.
The idea underlying this concept is that the transformations on the main
memory of strict load/store instruction set architectures achievable by
applying threads to them are primarily determined by the address width,
the word length, the operating unit size, and the cardinality of their
instruction set and the number of states of the threads that can be
applied to them.

We choose to use Maurer machines and basic thread algebra to study issues
relevant to the design of instruction set architectures.
Maurer machines are based on the view that a computer has a memory, the
contents of all memory elements make up the state of the computer, the
computer processes instructions, and the processing of an instruction
amounts to performing an operation on the state of the computer which
results in changes of the contents of certain memory elements.
The design of instruction set architectures must deal with these aspects
of real computers.
Turing machines and the other kinds of machines known from theoretical
computer science (see e.g.~\cite{HMU01a}) abstract from these aspects of
real computers.
Basic thread algebra is a form of process algebra.
Well-known process algebras, such as ACP~\cite{BW90},
CCS~\cite{Mil89}, and CSP~\cite{Hoa85}, are too general for our
purpose, viz.\ modelling deterministic sequential processes that
interact with a machine.
Basic thread algebra has been designed as an algebra of processes of
this kind.
In~\cite{BM05c}, we show that the processes considered in basic thread
algebra can be viewed as processes that are definable over ACP.
However, their modelling and analysis is rather difficult using such a
general process algebra.

The structure of this paper is as follows.
First, we review basic thread algebra (Section~\ref{sect-BTA}), Maurer
machines (Section~\ref{sect-MM}) and the operators for applying threads
to Maurer machines (Section~\ref{sect-apply}).
Next, we introduce the concept of a strict load/store Maurer instruction
set architecture (Section~\ref{sect-ISA}).
Then, we study the consequences of reducing the operating unit size of a
strict load/store Maurer instruction set architecture
(Section~\ref{sect-reducing-size}).
After that, we give conditions under which all possible transformations
on the states of the main memory of a strict load/store Maurer ISA with
a certain address width and word length can be achieved by applying a
thread to such a strict load/store Maurer ISA
(Section~\ref{sect-TPF-complete}).
Following this, we give a condition under which not all possible
transformations can be achieved (Section~\ref{sect-TPF-incomplete}).
Finally, we make some concluding remarks (Section~\ref{sect-conclusions}).

\section{Basic Thread Algebra}
\label{sect-BTA}

In this section, we review \BTA\ (Basic Thread Algebra), a form of
process algebra which was first presented in~\cite{BL02a} under the name
\BPPA\ (Basic Polarized Process Algebra).
It is a form of process algebra which is tailored to the description of
the behaviour of deterministic sequential programs under execution.
The behaviours concerned are called \emph{threads}.

In \BTA, it is assumed that there is a fixed but arbitrary set of
\emph{basic actions} $\BAct$.
\BTA\ has the following constants and operators:
\begin{itemize}
\item
the \emph{deadlock} constant $\Dead$;
\item
the \emph{termination} constant $\Stop$;
\item
for each $a \in \BAct$, a binary \emph{postconditional composition}
operator $\pcc{\ph}{a}{\ph}\,$.
\end{itemize}
We use infix notation for postconditional composition.
We introduce \emph{action prefixing} as an abbreviation:
$a \bapf p$, where $p$ is a term of \BTA, abbreviates
$\pcc{p}{a}{p}$.

The intuition is that each basic action performed by a thread is taken
as a command to be processed by the execution environment of the thread.
The processing of a command may involve a change of state of the
execution environment.
At completion of the processing of the command, the execution
environment produces a reply.
This reply is either $\True$ or $\False$ and is returned to the thread
concerned.
Let $p$ and $q$ be closed terms of \BTA.
Then $\pcc{p}{a}{q}$ will perform action $a$, and after that proceed
as $p$ if the processing of $a$ leads to the reply $\True$ (called a
positive reply) and proceed as $q$ if the processing of $a$ leads to
the reply $\False$ (called a negative reply).

Each closed term of \BTA\ denotes a finite thread, i.e.\ a thread whose
length of the sequences of actions that it can perform is bounded.
Guarded recursive specifications give rise to infinite threads.

A \emph{guarded recursive specification} over \BTA\ is a set of
recursion equations $E = \set{X = t_X \where X \in V}$, where $V$ is a
set of variables and each $t_X$ is a term of the form $\Dead$,
$\Stop$ or $\pcc{t}{a}{t'}$ with $t$ and $t'$ terms of \BTA\ that
contain only variables from $V$.
We write $\vars(E)$ for the set of all variables that occur on the
left-hand side of an equation in $E$.
We are only interested in models of \BTA\ in which guarded recursive
specifications have unique solutions, such as the projective limit model
of \BTA\ presented in~\cite{BB03a}.

We extend \BTA\ with guarded recursion by adding constants for solutions
of guarded recursive specifications and axioms concerning these
additional constants.
For each guarded recursive specification $E$ and each $X \in \vars(E)$,
we add a constant standing for the unique solution of $E$ for $X$ to the
constants of \BTA.
The constant standing for the unique solution of $E$ for $X$ is denoted
by $\rec{X}{E}$.
Moreover, we add the axioms for guarded recursion given in
Table~\ref{axioms-recursion} to \BTA,%
\begin{table}[!t]
\caption{Axioms for guarded recursion}
\label{axioms-recursion}
\begin{eqntbl}
\begin{saxcol}
\rec{X}{E} = \rec{t_X}{E} & \mif X \!=\! t_X \in E      & \axiom{RDP} \\
E \Implies X = \rec{X}{E} & \mif X \in \vars(E)         & \axiom{RSP}
\end{saxcol}
\end{eqntbl}
\end{table}
where we write $\rec{t_X}{E}$ for $t_X$ with, for all $Y \in \vars(E)$,
all occurrences of $Y$ in $t_X$ replaced by $\rec{Y}{E}$.
In this table, $X$, $t_X$ and $E$ stand for an arbitrary variable, an
arbitrary term of \BTA\ and an arbitrary guarded recursive
specification, respectively.
Side conditions are added to restrict the variables, terms and guarded
recursive specifications for which $X$, $t_X$ and $E$ stand.
The additional axioms for guarded recursion are known as the recursive
\pagebreak[2]
definition principle (RDP) and the recursive specification principle
(RSP).
The equations $\rec{X}{E} = \rec{t_X}{E}$ for a fixed $E$ express that
the constants $\rec{X}{E}$ make up a solution of $E$.
The conditional equations $E \Implies X = \rec{X}{E}$ express that this
solution is the only one.

We will write \BTA+\REC\ for \BTA\ extended with the constants for
solutions of guarded recursive specifications and axioms RDP and RSP.

Closed terms of \BTA+\REC\ that denote the same infinite thread cannot
always be proved equal by means of the axioms of \BTA+\REC.
We introduce the approximation induction principle to remedy this.
The approximation induction principle, \AIP\ in short, is based on the
view that two threads are identical if their approximations up to any
finite depth are identical.
The approximation up to depth $n$ of a thread is obtained by cutting it
off after performing a sequence of actions of length $n$.

\AIP\ is the infinitary conditional equation given in
Table~\ref{axioms-AIP}.%
\begin{table}[!t]
\caption{Approximation induction principle}
\label{axioms-AIP}
\begin{eqntbl}
\begin{axcol}
\AND{n \geq 0} \proj{n}{x} = \proj{n}{y} \Implies x = y  & \axiom{AIP}
\end{axcol}
\end{eqntbl}
\end{table}
Here, following~\cite{BB03a}, approximation of depth $n$ is phrased in
terms of a unary \emph{projection} operator $\proj{n}{\ph}$.
The axioms for the projection operators are given in
Table~\ref{axioms-projection}.%
\begin{table}[!t]
\caption{Axioms for projection operators}
\label{axioms-projection}
\begin{eqntbl}
\begin{axcol}
\proj{0}{x} = \Dead                                      & \axiom{P0} \\
\proj{n+1}{\Stop} = \Stop                                & \axiom{P1} \\
\proj{n+1}{\Dead} = \Dead                                & \axiom{P2} \\
\proj{n+1}{\pcc{x}{a}{y}} =
                       \pcc{\proj{n}{x}}{a}{\proj{n}{y}} & \axiom{P3}
\end{axcol}
\end{eqntbl}
\end{table}
In this table, $a$ stands for an arbitrary member of $\BAct$.

Henceforth, we write $\Efin(A)$, where $A \subseteq \BAct$, for the set
of all finite guarded recursive specifications over \BTA\ that contain
only postconditional operators $\pcc{\ph}{a}{\ph}$ for which
$a \in A$.
Moreover, we write $\Tfin(A)$, where $A \subseteq \BAct$, for the set of
all closed terms of \BTA+\REC\ that contain only postconditional
operators $\pcc{\ph}{a}{\ph}$ for which $a \in A$ and only constants
$\rec{X}{E}$ for which $E \in \Efin(A)$.

A \emph{linear recursive specification} over \BTA\ is a guarded
recursive specification $E = \set{X = t_X \where X \in V}$, where each
$t_X$ is a term of the form $\Dead$, $\Stop$ or $\pcc{Y}{a}{Z}$
with $Y,Z \in V$.
For each closed term $p \in \Tfin(A)$, there exist a linear recursive
specification $E \in \Efin(A)$ and a variable $X \in \vars(E)$ such that
$p = \rec{X}{E}$ is derivable from the axioms of \BTA+\REC.

Henceforth, we write $\Efinlin(A)$, where $A \subseteq \BAct$, for the set
of all linear recursive specifications from $\Efin(A)$.

Below, the interpretations of the constants and operators of \BTA+\REC\
in models of \BTA+\REC\ are denoted by the constants and operators
themselves.
Let $\cM$ be some model of \BTA+\REC, and let $p$ be an element from the
domain of $\cM$.
Then the set of \emph{states} or \emph{residual threads} of $p$,
written $\Res(p)$, is inductively defined as follows:
\begin{iteml}
\item
$p \in \Res(p)$;
\item
if $\pcc{q}{a}{r} \in \Res(p)$, then $q \in \Res(p)$ and
$r \in \Res(p)$.
\end{iteml}
We are only interested in models of \BTA+\REC\ in which
$\card(\Res(\rec{X}{E})) \leq \card(E)$ for all finite linear recursive
specifications $E$, such as the projective limit model of \BTA\ presented
in~\cite{BB03a}.

\section{Maurer Machines}
\label{sect-MM}

In this section, we review the concept of a Maurer machine.
This concept was first introduced in~\cite{BM05f}.

A \emph{Maurer machine} $H$ consists of the following components:
\begin{iteml}
\item
a non-empty set $M$;
\item
a set $B$ with $\card(B) \geq 2$;
\item
a set $\cS$ of functions $\funct{S}{M}{B}$;
\item
a set $\cO$ of functions $\funct{O}{\cS}{\cS}$;
\item
a set $A \subseteq \BAct$;
\item
a function $\funct{\Eval{\ph}}{A}{(\cO \x M)}$;
\end{iteml}
and satisfies the following conditions:
\begin{iteml}
\item
if $S_1,S_2 \in \cS$, $M' \subseteq M$, and $\funct{S_3}{M}{B}$ is
such that $S_3(x) = S_1(x)$ if $x \in M'$ and $S_3(x) = S_2(x)$ if
$x \not\in M'$, then $S_3 \in \cS$;
\item
if $S_1,S_2 \in \cS$, then the set
$\set{x \in M \where S_1(x) \neq S_2(x)}$ is finite;
\item
if $S \in \cS$, $a \in A$, and $\Eval{a} = \tup{O,m}$, then
$S(m) \in \set{\True,\False}$.
\end{iteml}
$M$ is called the \emph{memory} of $H$,
$B$ is called the \emph{base set} of $H$,
the members of $\cS$ are called the \emph{states} of $H$,
the members of $\cO$ are called the \emph{operations} of $H$,
the members of $A$ are called the \emph{basic actions} of $H$, and
$\Eval{\ph}$ is called the \emph{basic action interpretation function}
of $H$.

We write $M_H$, $B_H$, $\cS_H$, $\cO_H$, $A_H$ and $\Eval{\ph}_H$, where
$H = \tup{M,B,\cS,\cO,A,\Eval{\ph}}$ is a Maurer machine, for $M$, $B$,
$\cS$, $\cO$, $A$ and $\Eval{\ph}$, respectively.

A Maurer machine has much in common with a real computer.
The memory of a Maurer machine consists of memory elements whose
contents are elements from its base set.
The term memory must not be taken too strict.
For example, register files and caches must be regarded as parts of the
memory.
The contents of all memory elements together make up a state of the
Maurer machine.
State changes are accomplished by performing its operations.
Every state change amounts to changes of the contents of certain memory
elements.
The basic actions of a Maurer machine are the instructions that it is
able to process.
The processing of a basic action amounts to performing the operation
associated with the basic action by the basic action interpretation
function.
At completion of the processing, the content of the memory element
associated with the basic action by the basic action interpretation
function is the reply produced by the Maurer machine.
The term basic action originates from \BTA.

The first condition on the states of a Maurer machine is a structural
condition and the second one is a finite variability condition.
We return to these conditions, which are met by any real computer, after
the introduction of the input region and output region of an operation.
The third condition on the states of a Maurer machine restricts the
possible replies at completion of the processing of a basic action to
$\True$ and $\False$.

In~\cite{Mau66a}, a model for computers is proposed.
In~\cite{BM05f}, we introduced the term Maurer computer for what is a
computer according to Maurer's definition.
Leaving out the set of basic actions and the basic action interpretation
function from a Maurer machine yields a Maurer computer.
The set of basic actions and the basic action interpretation function
constitute the interface of a Maurer machine with its environment, which
effectuates state changes by issuing basic actions.

The notions of input region of an operation and output region of an
operation, which originate from~\cite{Mau66a}, are used in subsequent
sections.

Let $H = \tup{M,B,\cS,\cO,A,\Eval{\ph}}$ be a Maurer machine, and let
$\funct{O}{\cS}{\cS}$.
Then the \emph{input region} of $O$, written $\ireg(O)$, and the
\emph{output region} of $O$, written $\oreg(O)$, are the subsets of $M$
defined as follows:
\begin{ldispl}
\ireg(O) =
\bset{x \in M \bwhere
      \Exists{S_1,S_2 \in \cS}{{}}
       (\Forall{z \in M \diff \set{x}}{S_1(z) = S_2(z)} \And {}
\\
\phantom{\ireg(O) = \{x \in M \where \Exists{S_1,S_2 \in \cS}{(}}
       \Exists{y \in \oreg(O)}{O(S_1)(y) \neq O(S_2)(y)})}\;,
\eqnsep
\oreg(O) =
\bset{x \in M \bwhere
      \Exists{S \in \cS}{S(x) \neq O(S)(x)}}\;.\footnotemark
\end{ldispl}%
\footnotetext
{The following precedence conventions are used in logical formulas.
 Operators bind stronger than predicate symbols, and predicate symbols
 bind stronger than logical connectives and quantifiers.
 Moreover, $\Not$ binds stronger than $\And$ and $\Or$, and $\And$ and
 $\Or$ bind stronger than $\Implies$ and $\Iff$.
 Quantifiers are given the smallest possible scope.}%
$\oreg(O)$ is the set of all memory elements that are possibly affected
by $O$; and $\ireg(O)$ is the set of all memory elements that possibly
affect elements of $\oreg(O)$ under $O$.
For example, the input region and output region of an operation that
adds the content of a given main memory cell, say $X$, to the content of
a given register, say $R0$, are $\set{X,R0}$ and $\set{R0}$,
respectively.

Let $H = \tup{M,B,\cS,\cO,A,\Eval{\ph}}$ be a Maurer machine, let
$S_1,S_2 \in \cS$, and let $O \in \cO$.
Then $S_1 \restr \ireg(O) = S_2 \restr \ireg(O)$ implies
$O(S_1) \restr \oreg(O) = O(S_2) \restr \oreg(O)$.%
\footnote
{We use the notation $f \restr D$, where $f$ is a function and
 $D \subseteq \dom(f)$, for the function $g$ with $\dom(g) = D$ such
 that for all $d \in \dom(g)$,\, $g(d) = f(d)$.}
In other words, every operation transforms states that coincide on the
input region of the operation to states that coincide on the output
region of the operation.
The second condition on the states of a Maurer machine is necessary for
this fundamental property to hold.
The first condition on the states of a Maurer machine could be relaxed
somewhat.

In~\cite{Mau66a}, more results relating to input regions and output
regions are given.
Recently, a revised and expanded version of~\cite{Mau66a}, which
includes all the proofs, has appeared in~\cite{Mau06a}.

\section{Applying Threads to Maurer Machines}
\label{sect-apply}

In this section, we add for each Maurer machine $H$ a binary
\emph{apply} operator $\apply{\ph}{H}{\ph}$ to BTA+\REC\ and introduce
a notion of computation in the resulting setting.

The apply operators associated with Maurer machines are related to the
apply operators introduced in~\cite{BP02a}.
They allow for threads to transform states of the associated Maurer
machine by means of its operations.
Such state transformations produce either a state of the associated
Maurer machine or the \emph{undefined state} $\undef$.
It is assumed that $\undef$ is not a state of any Maurer machine.
We extend function restriction to $\undef$ by stipulating that
$\undef \restr M = \undef$ for any set $M$.
The first operand of the apply operator $\apply{\ph}{H}{\ph}$
associated with Maurer machine $H = \tup{M,B,\cS,\cO,A,\Eval{\ph}}$
must be a term from $\Tfin(A)$ and its second argument must be a state
from $\cS \union \set{\undef}$.

Let $H = \tup{M,B,\cS,\cO,A,\Eval{\ph}}$ be a Maurer machine, let
$p \in \Tfin(A)$, and let $S \in \cS$.
Then $\apply{p}{H}{S}$ is the state that results if all basic actions
performed by thread $p$ are pro\-cessed by the Maurer machine $H$ from
initial state $S$.
The processing of a basic action $a$ by $H$ amounts to a state
change according to the operation associated with $a$ by $\Eval{\ph}$.
In the resulting state, the reply produced by $H$ is contained in the
memory element associated with $a$ by $\Eval{\ph}$.
If $p$ is $\Stop$, then there will be no state change.
If $p$ is $\Dead$, then the result is $\undef$.

Let $H = \tup{M,B,\cS,\cO,A,\Eval{\ph}}$ be a Maurer machine, and let
$\tup{O_a,m_a} = \Eval{a}$ for all $a \in A$.
Then the apply operator $\apply{\ph}{H}{\ph}$ is defined by the
equations given in Table~\ref{axioms-apply}%
\begin{table}[!t]
\caption{Defining equations for apply operator}
\label{axioms-apply}
\begin{eqntbl}
\begin{seqncol}
\apply{x}{H}{\undef} = \undef \\
\apply{\Stop}{H}{S} = S \\
\apply{\Dead}{H}{S} = \undef \\
\apply{(\pcc{x}{a}{y})}{H}{S} = \apply{x}{H}{O_a(S)}
 & \mif O_a(S)(m_a) = \True \\
\apply{(\pcc{x}{a}{y})}{H}{S} = \apply{y}{H}{O_a(S)}
 & \mif O_a(S)(m_a) = \False
\end{seqncol}
\end{eqntbl}
\end{table}
and the rule given in Table~\ref{rule-apply}.%
\begin{table}[!t]
\caption{Rule for divergence}
\label{rule-apply}
\begin{eqntbl}
\begin{seqncol}
\AND{n \geq 0} \apply{\proj{n}{x}}{H}{S} = \undef \Implies
\apply{x}{H}{S} = \undef
\end{seqncol}
\end{eqntbl}
\end{table}
In these tables, $a$ stands for an arbitrary member of $A$ and $S$
stands for an arbitrary member of $\cS$.

Let $H = \tup{M,B,\cS,\cO,A,\Eval{\ph}}$ be a Maurer machine, let
$p \in \Tfin(A)$, and let $S \in \cS$.
Then $p$ \emph{converges} from $S$ on $H$ if there exists an
$n \in \Nat$ such that $\apply{\proj{n}{p}}{H}{S} \neq \undef$.
The rule from Table~\ref{rule-apply} can be read as follows: if $x$ does
not converge from $S$ on $H$, then $\apply{x}{H}{S}$ equals $\undef$.

Below, we introduce a notion of computation in the current setting.
First, we introduce some auxiliary notions.

Let $H = \tup{M,B,\cS,\cO,A,\Eval{\ph}}$ be a Maurer machine, and let
$\tup{O_a,m_a} = \Eval{a}$ for all $a \in A$.
Then the \emph{step relation}
\pagebreak[2]
$\step{\ph}{H}{\ph} \subseteq (\Tfin(A) \x \cS) \x (\Tfin(A) \x \cS)$
is inductively defined as follows:
\begin{iteml}
\item
if $O_a(S)(m_a) = \True$  and $p = \pcc{p'}{a}{p''}$, then
$\step{\tup{p,S}}{H}{\tup{p',O_a(S)}}$;
\item
if $O_a(S)(m_a) = \False$ and $p = \pcc{p'}{a}{p''}$, then
$\step{\tup{p,S}}{H}{\tup{p'',O_a(S)}}$.
\end{iteml}

Let $H = \tup{M,B,\cS,\cO,A,\Eval{\ph}}$ be a Maurer machine.
Then a \emph{full path} in $\step{\ph}{H}{\ph}$ is one of the
following:
\begin{iteml}
\item
a finite path $\seq{\tup{p_0,S_0},\ldots,\tup{p_n,S_n}}$ in
$\step{\ph}{H}{\ph}$ such that there exists no
$\tup{p_{n+1},S_{n+1}} \in \Tfin(A) \x \cS$ with
$\step{\tup{p_n,S_n}}{H}{\tup{p_{n+1},S_{n+1}}}$;
\item
an infinite path $\seq{\tup{p_0,S_0},\tup{p_1,S_1},\ldots}$ in
$\step{\ph}{H}{\ph}$.
\end{iteml}
Moreover, let $p \in \Tfin(A)$, and let $S \in \cS$.
Then the \emph{full path} of $\conf{p}{H}{S}$ on $H$ is the unique
full path in $\step{\ph}{H}{\ph}$ from $\tup{p,S}$.
If $p$ converges from $S$ on $H$, then the full path of
$\conf{p}{H}{S}$ on $H$ is called the \emph{computation} of
$\conf{p}{H}{S}$ on $H$ and we write $\steps{\conf{p}{H}{S}}{H}$
for the length of the computation of $\conf{p}{H}{S}$ on $H$.

It is easy to see that
$\step{\tup{p_0,S_0}}{H}{\tup{p_1,S_1}}$ only if
$\apply{p_0}{H}{S_0} = \apply{p_1}{H}{S_1}$ and that
$\seq{\tup{p_0,S_0},\ldots,\tup{p_n,S_n}}$ is the computation of
$\conf{p_0}{H}{S_0}$ on $H$ only if
$p_n = \Stop$ and $S_n = \apply{p_0}{H}{S_0}$.
It is also easy to see that, if $p_0$ converges from $S_0$ on $H$,
$\steps{\conf{p_0}{H}{S_0}}{H}$ is the least $n \in \Nat$ such that
$\apply{\proj{n}{p_0}}{H}{S_0} \neq \undef$.

\section{Instruction Set Architectures}
\label{sect-ISA}

In this section, we introduce the concept of a strict load/store Maurer
instruction set architecture.
This concept, which was first introduced in~\cite{BM06b}, takes its name
from the following: it is described in the setting of Maurer machines,
it concerns only load/store architectures, and the load/store
architectures concerned are strict in some respects that will be
explained after its formalization.

The concept of a strict load/store Maurer instruction set architecture,
or shortly a strict load/store Maurer ISA, is an approximation of the
concept of a load/store instruction set architecture
(see e.g.~\cite{HP03a}).
It is focussed on instructions for data manipulation and data transfer.
Transfer of program control is treated in a uniform way over different
strict load/store Maurer ISAs by working at the abstraction level of
threads.
All that is left of transfer of program control at this level is
postconditional composition.

Each Maurer machine has a number of basic actions with which an
operation is associated.
Henceforth, when speaking about Maurer machines that are strict
load/store Maurer ISAs, such basic actions are loosely called basic
instructions.
The term basic action is uncommon when we are concerned with ISAs.

The idea underlying the concept of a strict load/store Maurer ISA is
that there is a main memory whose elements contain data, an operating
unit with a small internal memory by which data can be manipulated, and
an interface between the main memory and the operating unit for data
transfer between them.
For the sake of simplicity, data is restricted to the natural numbers
between $0$ and some upper bound.
Other types of data that could be supported can always be represented
by the natural numbers provided.
Moreover, the data manipulation instructions offered by a strict
load/store Maurer ISA are not restricted and may include ones that are
tailored to manipulation of representations of other types of data.
Therefore, we believe that nothing essential is lost by the restriction
to natural numbers.

The concept of a strict load/store Maurer ISA is parametrized by:
\begin{iteml}
\item
an address width $\aw$;
\item
a word length $\wl$;
\item
an operating unit size $\ous$;
\item
a number $\nrpl$ of pairs of data and address registers for load
instructions;
\item
a number $\nrps$ of pairs of data and address registers for store
instructions;
\item
a set $\Adm$ of basic instructions for data manipulation;
\end{iteml}
where $\aw,\ous \geq 0$, $\wl,\nrpl,\nrps > 0$ and
$\Adm \subseteq \BAct$.

The address width $\aw$ can be regarded as the number of bits used for
the binary representation of addresses of data memory elements.
The word length $\wl$ can be regarded as the number of bits used to
represent data in data memory elements.
The operating unit size $\ous$ can be regarded as the number of bits
that the internal memory of the operating unit contains.
The operating unit size is measured in bits because this allows for
establishing results in which no assumption about the internal structure
of the operating unit are involved.

It is assumed that, for each $n \in \Nat$, a fixed but arbitrary
countably infinite set $\MData{n}$ and a fixed but arbitrary bijection
$\funct{\mdata{n}}{\Nat}{\MData{n}}$ have been given.
The members of $\MData{n}$ are called \emph{data memory elements}.
The contents of data memory elements are taken as data.
The data memory elements from $\MData{n}$ can contain natural numbers in
the interval $[0,2^n - 1]$.

It is assumed that a fixed but arbitrary countably infinite set $\MOU$
and a fixed but arbitrary bijection $\funct{\mou}{\Nat}{\MOU}$ have been
given.
The members of $\MOU$ are called \emph{operating unit memory elements}.
They can contain natural numbers in the set $\set{0,1}$, i.e.\ bits.
Usually, a part of the operating unit memory is partitioned into groups
to which data manipulation instructions can refer.

It is assumed that, for each $n \in \Nat$, fixed but arbitrary countably
infinite sets $\Mld{n}$, $\Mla{n}$, $\Msd{n}$ and $\Msa{n}$ and fixed but
arbitrary bijections $\funct{\mld{n}}{\Nat}{\Mld{n}}$,
$\funct{\mla{n}}{\Nat}{\Mla{n}}$, $\funct{\msd{n}}{\Nat}{\Msd{n}}$ and
$\funct{\msa{n}}{\Nat}{\Msa{n}}$ have been given.
The members of $\Mld{n}$, $\Mla{n}$, $\Msd{n}$ and $\Msa{n}$ are called
\emph{load data registers}, \emph{load address registers},
\emph{store data registers} and \emph{store address registers},
respectively.
The contents of load data registers and store data registers are taken
as data, whereas the contents of load address registers and store
address registers are taken as addresses.
The load data registers from $\Mld{n}$, the load address registers from
$\Mla{n}$, the store data registers from $\Msd{n}$ and the store address
registers from $\Msa{n}$ can contain natural numbers in the interval
$[0,2^n - 1]$.
The load and store registers are special memory elements designated for
transferring data between the data memory and the operating unit memory.

A single special memory element $\replr$ is taken for passing on the
replies resulting from the processing of basic instructions.
This special memory element is called the \emph{reply register}.

It is assumed that, for each $n,n' \in \Nat$, $\MData{n}$, $\MOU$,
$\Mld{n}$, $\Mla{n'}$, $\Msd{n}$, $\Msa{n'}$ and $\set{\replr}$ are
pairwise disjoint sets.

If $M \subseteq \MData{n}$ and $\mdata{n}(i) \in M$, then we write
$M[i]$ for $\mdata{n}(i)$.
If $M \subseteq \Mld{n}$ and $\mld{n}(i) \in M$, then we write $M[i]$
for $\mld{n}(i)$.
If $M \subseteq \Mla{n}$ and $\mla{n}(i) \in M$, then we write $M[i]$
for $\mla{n}(i)$.
If $M \subseteq \Msd{n}$ and $\msd{n}(i) \in M$, then we write $M[i]$
for $\msd{n}(i)$.
If $M \subseteq \Msa{n}$ and $\msa{n}(i) \in M$, then we write $M[i]$
for $\msa{n}(i)$.
Moreover, we write $\Bool$ for the set $\set{\True,\False}$.

Let $\aw,\ous \geq 0$, $\wl,\nrpl,\nrps > 0$ and $\Adm \subseteq \BAct$.
Then a \emph{strict load/store Maurer instruction set architecture} with
parameters $\aw$, $\wl$, $\ous$, $\nrpl$, $\nrps$ and $\Adm$ is a
Maurer machine $H = \tup{M,B,\cS,\cO,A,\Eval{\ph}}$ with
\begin{ldispl}
\begin{aeqns}
M & = &
\MMData \union \MMOU \union
\MMld \union \MMla \union \MMsd \union \MMsa \union
\set{\replr}\;,
\\
B & = & [0,2^\wl-1] \union [0,2^\aw-1] \union \Bool\;,
\\
\cS & = &
\set{\funct{S}{M}{B} \where {}
\\ & & \phantom{\{}
     \Forall{m \in \MMData \union
                   \MMld \union \MMsd}
      {S(m) \in [0,2^\wl-1]} \And {}
\\ & & \phantom{\{}
     \Forall{m \in \MMla \union \MMsa}
      {S(m) \in [0,2^\aw-1]} \And {}
\\ & & \phantom{\{}
     \Forall{m \in \MMOU}{S(m) \in \set{0,1}} \And
     S(\replr) \in \Bool}\;,
\\
\cO & = & \set{O_a \where a \in A}\;,
\\
A & = &
\set{\load{:}n \where n \in [0,\nrpl-1]} \union
\set{\store{:}n \where n \in [0,\nrps-1]} \union \Adm\;,
\\
\Eval{a} & = & \tup{O_a,\replr} \quad \mathrm{for\; all}\; a \in A\;,
\end{aeqns}
\end{ldispl}%
where
\begin{ldispl}
\begin{aeqns}
\MMData & = & \set{\mdata{\wl}(i) \where i \in [0,2^\aw-1]}\;, \\
\MMOU   & = & \set{\mou(i) \where i \in [0,\ous-1]}\;, \\
\MMld   & = & \set{\mld{\wl}(i) \where i \in [0,\nrpl-1]}\;, \\
\MMla   & = & \set{\mla{\aw}(i) \where i \in [0,\nrpl-1]}\;, \\
\MMsd   & = & \set{\msd{\wl}(i) \where i \in [0,\nrps-1]}\;, \\
\MMsa   & = & \set{\msa{\aw}(i) \where i \in [0,\nrps-1]}\;,
\end{aeqns}
\end{ldispl}%
and, for all $n \in [0,\nrpl-1]$, $O_{\load{:}n}$ is the unique
function from $\cS$ to $\cS$ such that for all $S \in \cS$:
\begin{ldispl}
\begin{aeqns}
O_{\load{:}n}(S) \restr (M \diff \set{\MMld[n],\replr}) & = &
S \restr (M \diff \set{\MMld[n],\replr})\;,
\\
O_{\load{:}n}(S)(\MMld[n]) & = & S(\MMData[S(\MMla[n])])\;,
\\
O_{\load{:}n}(S)(\replr) & = & \True\;,
\end{aeqns}
\end{ldispl}%
and, for all $n \in [0,\nrps-1]$, $O_{\store{:}n}$ is the unique
function from $\cS$ to $\cS$ such that for all $S \in \cS$:
\begin{ldispl}
\begin{aeqns}
O_{\store{:}n}(S) \restr
 (M \diff \set{\MMData[S(\MMsa[n])],\replr}) & = &
\\
\multicolumn{3}{@{}r@{}}
 {S \restr
  (M \diff \set{\MMData[S(\MMsa[n])],\replr})\;,}
\\
O_{\store{:}n}(S)(\MMData[S(\MMsa[n])]) & = & S(\MMsd[n])\;,
\\
O_{\store{:}n}(S)(\replr) & = & \True\;,
\end{aeqns}
\end{ldispl}%
and, for all $a \in \Adm$,\, $O_a$ is a function from $\cS$ to $\cS$
such that:
\begin{ldispl}
\begin{aeqns}
\ireg(O_a) & \subseteq & \MMOU \union \MMld\;,
\eqnsep
\oreg(O_a) & \subseteq &
\MMOU \union \MMla \union \MMsd \union \MMsa \union
\set{\replr}\;.
\end{aeqns}
\end{ldispl}%
We will write $\MISA(\aw,\wl,\ous,\nrpl,\nrps,\Adm)$ for the set of all
strict load/store Maurer ISAs with parameters $\aw$, $\wl$, $\ous$,
$\nrpl$, $\nrps$ and $\Adm$.

In our opinion, load/store architectures give rise to a relatively
simple interface between the data memory and the operating unit.

A strict load/store Maurer ISA is strict in the following respects:
\begin{iteml}
\item
with data transfer between the data memory and the operating unit, a
strict separation is made between data registers for loading, address
registers for loading, data registers for storing, and address registers
for storing;
\item
from these registers, only the registers of the first kind are allowed
in the input regions of data manipulation operations, and only the
registers of the other three kinds are allowed in the output regions of
data manipulation operations;
\item
a data memory whose size is less than the number of addresses determined
by the address width is not allowed.
\end{iteml}
The first two ways in which a strict load/store Maurer ISA is strict
concern the interface between the data memory and the operating unit.
We believe that they yield the most conveniently arranged interface for
theoretical work relevant to the design of instruction set
architectures.
The third way in which a strict load/store Maurer ISA is strict saves
the need to deal with addresses that do not address a memory element.
Such addresses can be dealt with in many different ways, each of which
complicates the architecture considerably.
We consider their exclusion desirable in much theoretical work relevant
to the design of instruction set architectures.

A strict separation between data registers for loading, address
registers for loading, data registers for storing, and address registers
for storing is also made in Cray and Thornton's design of the CDC~6600
computer~\cite{Tho70a}, which is arguably the first implemented
load/store architecture.
However, in their design, data registers for loading are also allowed in
the input regions of data manipulation operations.

\section{Reducing the Operating Unit Size}
\label{sect-reducing-size}

In a strict load/store Maurer ISA, data manipulation takes place in the
operating unit.
This raises questions concerning the consequences of changing the
operating unit size.
One of the questions is whether, if the operating unit size is reduced
by one, it is possible with new instructions for data manipulation to
transform each thread that can be applied to the original ISA into one
or more threads that can each be applied to the ISA with the reduced
operating unit size and together yield the same state changes on the
data memory.
This question can be answered in the affirmative.

\begin{theorem}
\label{theorem-reducing-ous}
Let $\aw \geq 0$, $\wl,\ous,\nrpl,\nrps > 0$ and $\Adm \subseteq \BAct$,
let
$H = \tup{M,B,\cS,\cO,A,\Eval{\ph}} \in
 \MISA(\aw,\wl,\ous,\nrpl,\nrps,\Adm)$, and let
$\MMData = \set{\mdata{\wl}(i) \where i \in [0,2^\aw-1]}$ and
$\bc = \mou(\ous-1)$.
Then there exist an $\Admp \subseteq \BAct$ and an
$H' = \tup{M',B,\cS',\cO',A',\Eval{\ph}'} \in
 \MISA(\aw,\wl,\ous-1,\nrpl,\nrps,\Admp)$
such that for all $p \in \Tfin(A)$ there exist $p'_0,p'_1 \in \Tfin(A')$
such that
\begin{ldispl}
\set{\tup{S \restr \MMData,
          (\apply{p}{H}{S}) \restr \MMData}
     \where S \in \cS \And S(\bc) = 0}
\\ \quad {} =
\set{\tup{S' \restr \MMData,
          (\apply{p'_0}{H'}{S'}) \restr \MMData}
     \where S' \in \cS'}
\end{ldispl}%
and
\begin{ldispl}
\set{\tup{S \restr \MMData,
          (\apply{p}{H}{S}) \restr \MMData}
     \where S \in \cS \And S(\bc) = 1}
\\ \quad {} =
\set{\tup{S' \restr \MMData,
          (\apply{p'_1}{H'}{S'}) \restr \MMData}
     \where S' \in \cS'}\;.
\end{ldispl}%
\end{theorem}
Notice that $\bc$ is the operating unit memory element of $H$ that is
missing in $H'$.
In the proof of Theorem~\ref{theorem-reducing-ous} given below, we take
$\Admp$ such that, for each instruction $a$ in $\Adm$, there are four
instructions $a(0)$, $a(1)$, $\ol{a}(0)$ and $\ol{a}(1)$ in $\Admp$.
$O_{a(0)}$ and $O_{a(1)}$ affect the memory elements of $H'$ like $O_a$
would affect them if the content of the missing operating unit memory
element would be $0$ and $1$, respectively.
The effect that $O_a$ would have on the missing operating unit memory
element is made available by $O_{\ol{a}(0)}$ and $O_{\ol{a}(1)}$,
respectively.
They do nothing but replying $\False$ if the content of the missing
operating unit memory element would become $0$ and $\True$ if the
content of the missing operating unit memory element would become $1$.
\begin{proof}[of Theorem~\ref{theorem-reducing-ous}]
Instead of the result to be proved, we prove that there exist an
$\Admp \subseteq \BAct$ and an
$H' = \tup{M',B,\cS',\cO',A',\Eval{\ph}'} \in
 \MISA(\aw,\wl,\ous-1,\linebreak\nrpl,\nrps,\Admp)$
such that for all $p \in \Tfin(A)$ there exist $p'_0,p'_1 \in \Tfin(A')$
such that
\begin{ldispl}
\set{\tup{S \restr (M' \diff \set{\replr}),(\apply{p}{H}{S}) \restr (M' \diff \set{\replr})}
     \where S \in \cS \And S(\bc) = 0}
\\ \quad  {} =
\set{\tup{S' \restr (M' \diff \set{\replr}),(\apply{p'_0}{H'}{S'}) \restr (M' \diff \set{\replr})}
     \where S' \in \cS'}
\end{ldispl}%
and
\begin{ldispl}
\set{\tup{S \restr (M' \diff \set{\replr}),(\apply{p}{H}{S}) \restr (M' \diff \set{\replr})}
     \where S \in \cS \And S(\bc) = 1}
\\ \quad  {} =
\set{\tup{S' \restr (M' \diff \set{\replr}),(\apply{p'_1}{H'}{S'}) \restr (M' \diff \set{\replr})}
     \where S' \in \cS'}\;.
\end{ldispl}%
This is sufficient because $\MMData \subseteq M' \diff \set{\replr}$.

We take
$\Admp = \set{a(k),\ol{a}(k) \where a \in \Adm \And k \in \set{0,1}}$,
and we take $H' = \tup{M',B,\cS',\cO',A',\Eval{\ph}'}$ such that, for
each $a \in \Adm$ and $k \in \set{0,1}$, $O_{a(k)}$ and $O_{\ol{a}(k)}$
are the unique functions from $\cS'$ to $\cS'$ such that for all
$S' \in \cS'$:
\begin{ldispl}
\begin{aeqns}
O_{a(k)}(S') & = & O_{a}(\rho_k(S')) \restr M'\;,
\eqnsep
O_{\ol{a}(k)}(S') \restr (M' \diff \set{\replr}) & = &
S' \restr (M' \diff \set{\replr})\;,
\\
O_{\ol{a}(k)}(S')(\replr) & = & \gamma(O_{a}(\rho_k(S'))(\bc))\;,
\end{aeqns}
\end{ldispl}%
where, for each $k \in \set{0,1}$, $\rho_k$ is the unique function from
$\cS'$ to $\cS$ such that
\begin{ldispl}
\begin{aeqns}
\rho_k(S') \restr M' & = & S'\;,
\\
\rho_k(S')(\bc)      & = & k
\end{aeqns}
\end{ldispl}%
and $\funct{\gamma}{\set{0,1}}{\Bool}$ is defined by
\begin{ldispl}
\begin{aeqns}
\gamma(0) & = & \False\;,
\\
\gamma(1) & = & \True\;.
\end{aeqns}
\end{ldispl}%

We restrict ourselves to
$p \in \set{\rec{X}{E}\where E \in \Efinlin(A) \And X \in \vars(E)}$,
because each term from $\Tfin(A)$ can be proved equal to some constant
from this set by means of the axioms of \BTA+\REC.

We define transformation functions
$\funct{\phi_k}
 {\set{\rec{X}{E}\where E \in \Efinlin(A) \And X \in \vars(E)}}
 {\set{\rec{X}{E}\where E \in \Efinlin(A') \And X \in \vars(E)}}$,
for $k \in \set{0,1}$, as follows:
\begin{ldispl}
\begin{aeqns}
\phi_k(\rec{X}{E}) = \rec{X_k}{\phi'_k(E)}\;,
\end{aeqns}
\end{ldispl}%
where $\funct{\phi'_k}{\Efinlin(A)}{\Efinlin(A')}$, for
$k \in \set{0,1}$ is defined as follows:
\begin{ldispl}
\begin{aeqns}
\phi'_k(\set{X = \Stop}) & = & \set{X_k = \Stop}\;,
\\
\phi'_k(\set{X = \Dead}) & = & \set{X_k = \Dead}\;,
\\
\phi'_k(\set{X = \pcc{Y}{a}{Z}}) & = &
\set{X_k = \pcc{Y_k}{a}{Z_k}} & \mif a \not\in \Adm\;,
\\
\phi'_k(\set{X = \pcc{Y}{a}{Z}}) & = &
\set{X_k = \pcc{X'_k}{\ol{a}(k)}{X''_k},
\\ & & \phantom{\{}
     X'_k = \pcc{Y_1}{a(k)}{Z_1},
\\ & & \phantom{\{}
     X''_k = \pcc{Y_0}{a(k)}{Z_0}} \quad & \mif a \in \Adm\;,
\\
\phi'_k(E' \union E'') & = & \phi'_k(E') \union \phi'_k(E'')\;.
\end{aeqns}
\end{ldispl}%
Here, for each variable $X$, the new variables $X_0$, $X'_0$, $X''_0$,
$X_1$, $X'_1$ and $X''_1$ are taken such that:
(i)~they are pairwise different variables;
(ii)~for each variable $Y$ different from $X$,
$\set{X_0,X'_0,X''_0,X_1,X'_1,X''_1}$ and
$\set{Y_0,Y'_0,Y''_0,Y_1,Y'_1,Y''_1}$ are disjoint sets.

Let $p \in \set{\rec{X}{E}\where E \in \Efinlin(A) \And X \in \vars(E)}$,
let $S \in \cS$ and $S' \in \cS'$ be such that $S \restr M' = S'$,
let $\tup{p_i,S_i}$ be the $(i+1)$st element in the full path of
$\tup{p,S}$ on $H$, and
let $\tup{p'_i,S'_i}$ be the $(i+1)$st element in the full path of
$\tup{\phi_{S(\bc)}(p),S'}$ on $H'$ whose first component does not
equal $\pcc{p}{a(k)}{q}$ for any $p,q \in \Tfin$, $a \in \Adm$ and
$k \in \set{0,1}$.
Moreover, let $a_i$ be the unique $a \in A$ such that
$p_i = \pcc{p}{a}{q}$ for some $p,q \in \Tfin$.
Then, it is easy to prove by induction on $i$ that if $a_i \in \Adm$:
\begin{eqnarray}
& &
O_{a_i}(S_i)(\bc) =
\gamma^{-1}(O_{\ol{a_i}(S_i(\bc))}(S'_i)(\replr))\;,
\label{eqn-reduce-ous-1}
\\
& &
O_{a_i}(S_i)(\replr)\hspace*{.3em} =
O_{{a_i}(S_i(\bc))}(O_{{\ol{a}_i}(S_i(\bc))}(S'_i))(\replr)
\label{eqn-reduce-ous-2}
\end{eqnarray}
(if $i+1 < \steps{\tup{p,S}}{H}$ in case $p$ converges from $S$ on $H$).
Now, using (\ref{eqn-reduce-ous-1}) and (\ref{eqn-reduce-ous-2}),
it is easy to prove by induction on $i$ that:
\begin{ldispl}
\phi_{S_i(\bc)}(p_i) = p'_i\;,
\\
S_i \restr (M \diff \set{\replr}) =
\rho_{S_i(\bc)}(S'_i) \restr (M \diff \set{\replr})
\end{ldispl}%
(if $i < \steps{\tup{p,S}}{H}$ in case $p$ converges from $S$ on $H$).
From this, the result follows immediately.
\end{proof}

The proof of Theorem~\ref{theorem-reducing-ous} gives us some upper
bounds:
\begin{iteml}
\item
for each thread that can be applied to the original ISA, the number of
threads that can together produce the same state changes on the data
memory of the ISA with the reduced operating unit does not have to be
more than $2$;
\item
the number of states of the new threads does not have to be more than
$6$ times the number of states of the original thread;
\item
the number of steps that the new threads take to produce some state
change does not have to be more than $2$ times the number of steps that
the original thread takes to produce that state change;
\item
the number of instructions of the ISA with the reduced operating unit
does not have to be more than $4$ times the number of instructions of
the original ISA.
\end{iteml}
Moreover, the proof indicates that more efficient new threads are
possible: equations $X = \pcc{Y}{a}{Z}$ with $a \in \Adm$ can be
treated as if $a \not\in \Adm$ in the case where the missing operating
unit memory element is not in $\ireg(O_a)$.

As a corollary of the proof of Theorem~\ref{theorem-reducing-ous}, we
have that only one transformed thread is needed if the input region of
the operation associated with the first instruction performed by the
original thread does not include the operating unit memory element that
is missing in the ISA with the reduced operating unit size.
\begin{corollary}
\label{corollary-reducing-ous-ireg}
Let $\aw \geq 0$, $\wl,\ous,\nrpl,\nrps > 0$ and $\Adm \subseteq \BAct$,
let
$H = \tup{M,B,\cS,\cO,A,\Eval{\ph}} \in
 \MISA(\aw,\wl,\ous,\nrpl,\nrps,\Adm)$, and let
$\MMData = \set{\mdata{\wl}(i) \where i \in [0,2^\aw-1]}$ and
$\bc = \mou(\ous-1)$.
Moreover, let
$\cT' =
 \set{\pcc{q}{a}{r} \where
      q,r \in \Tfin(A) \And a \in A \And \bc \in \ireg(O_a)}$.
Then there exist an $\Admp \subseteq \BAct$ and an
$H' = \tup{M',B,\cS',\cO',A',\Eval{\ph}'} \in
 \MISA(\aw,\wl,\ous-1,\nrpl,\nrps,\Admp)$
such that for all $p \in \Tfin(A) \diff \cT'$ there exists a
$p' \in \Tfin(A')$ such that
\begin{ldispl}
\set{\tup{S \restr \MMData,
          (\apply{p}{H}{S}) \restr \MMData}
     \where S \in \cS}
\\ \quad {} =
\set{\tup{S' \restr \MMData,
          (\apply{p'}{H'}{S'}) \restr \MMData}
     \where S' \in \cS'}\;.
\end{ldispl}%
\end{corollary}

As another corollary of the proof of Theorem~\ref{theorem-reducing-ous},
we have that, if the operating unit size is reduced to zero, it is still
possible to transform each thread that can be applied to the original
ISA into a number of threads that can each be applied to the ISA with
the reduced operating unit size and together yield the same state
changes on the data memory.
\begin{corollary}
\label{corollary-reducing-ous-zero}
Let $\aw \geq 0$, $\wl,\ous,\nrpl,\nrps > 0$ and $\Adm \subseteq \BAct$,
let
$H = \tup{M,B,\cS,\cO,A,\Eval{\ph}} \in
 \MISA(\aw,\wl,\ous,\nrpl,\nrps,\Adm)$, and let
$\MMData = \set{\mdata{\wl}(i) \where i \in [0,2^\aw-1]}$ and
$\MMOU = \set{\mou(i) \where i \in [0,\ous-1]}$.
Then there exist an $\Admp \subseteq \BAct$ and an
$H' = \tup{M',B,\cS',\cO',A',\Eval{\ph}'} \in
 \MISA(\aw,\wl,0,\nrpl,\linebreak[2]\nrps,\Admp)$
such that for all $p \in \Tfin(A)$ and
$S_\nm{ou} \in \set{S \restr \MMOU \where S \in \cS}$
there exists a $p' \in \Tfin(A')$ such that
\begin{ldispl}
\set{\tup{S \restr \MMData,
          (\apply{p}{H}{S}) \restr \MMData}
     \where S \in \cS \And S \restr \MMOU = S_\nm{ou}}
\\ \quad {} =
\set{\tup{S' \restr \MMData,
          (\apply{p'}{H'}{S'}) \restr \MMData}
     \where S' \in \cS'}\;.
\end{ldispl}%
\end{corollary}

Let $\MMOU$ be as in Corollary~\ref{corollary-reducing-ous-zero}.
Then the cardinality of $\set{S \restr \MMOU \where S \in \cS}$ is
$2^\ous$.
Therefore, we have that, if the operating unit size is reduced to zero,
at most $2^\ous$ transformed threads are needed for each thread that can
be applied to the original ISA.
Notice that Corollary~\ref{corollary-reducing-ous-zero} does not go
through if the number of states of the new threads is bounded.

\section{Thread Powered Function Classes}
\label{sect-TPF-complete}

A simple calculation shows that, for a strict load/store Maurer ISA with
address width $\aw$ and word length $\wl$, the number of possible
transformations on the states of the data memory is $2^{(2^{(2^\aw \cdot
\wl + \aw)} \cdot \wl)}$.
This raises questions concerning the possibility to achieve all these
state transformation by applying a thread to a strict load/store Maurer
ISA with this address width and word length.
One of the questions is how this possibi\-lity depends on the operating
unit size of the ISAs, the size of the instruction set of the ISAs, and
the maximal number of states of the threads.
This brings us to introduce the concept of a thread powered function
class.

The concept of a thread powered function class is parametrized by:
\begin{iteml}
\item
an address width $\aw$;
\item
a word length $\wl$;
\item
an operating unit size $\ous$;
\item
an instruction set size $\iss$;
\item
a state space bound $\ssb$;
\item
a working area flag $\waf$;
\end{iteml}
where $\aw,\ous \geq 0$, $\wl,\iss,\ssb > 0$ and $\waf \in \Bool$.

The instruction set size $\iss$ is the number of basic instructions,
excluding load and store instructions.
To simplify the setting, we consider only the case where there is one
load instruction and one store instruction.
The state space bound $\ssb$ is a bound on the number of states of the
thread that is applied.
The working area flag $\waf$ indicates whether a part of the data memory
is taken as a working area.
A part of the data memory is taken as a working area if we are not
interested in the state transformations with respect to that part.
To simplify the setting, we always set aside half of the data memory for
working area if a working area is in order.

Intuitively, the thread powered function class with parameters $\aw$,
$\wl$, $\ous$, $\iss$, $\ssb$ and $\waf$ are the transformations on the
states of the data memory or the first half of the data memory,
depending on $\waf$, that can be achieved by applying threads with not
more than $\ssb$ states to a strict load/store Maurer ISA of which the
address width is $\aw$, the word length is $\wl$, the operating unit
size is $\ous$, the number of register pairs for load instructions is
$1$, the number of register pairs for store instructions is $1$, and the
cardinality of the set of instructions for data manipulation is $\iss$.
Henceforth, we will use the term \emph{external memory} for the data
memory if $\waf = \False$ and for the first half of the data memory if
$\waf = \True$.
Moreover, if $\waf = \True$, we will use the term \emph{internal memory}
for the second half of the data memory.

For $\aw \geq 0$ and $\wl > 0$, we define $\Mext{\aw,\wl}$,
$\Sext{\aw,\wl}$ and $\SText{\aw,\wl}$ as follows:
\begin{ldispl}
\begin{aeqns}
\Mext{\aw,\wl}  & = &
\set{\mdata{\wl}(i) \where i \in [0,2^\aw-1]}\;,
\\
\Sext{\aw,\wl}  & = &
\set{S \where \funct{S}{\Mext{\aw,\wl}}{[0,2^\wl-1]}}\;,
\\
\SText{\aw,\wl} & = &
\set{T \where \funct{T}{\Sext{\aw,\wl}}{\Sext{\aw,\wl}}}\;.
\end{aeqns}
\end{ldispl}%
$\Mext{\aw,\wl}$ is the data memory of a strict load/store Maurer ISA
with address width $\aw$ and word length $\wl$,
$\Sext{\aw,\wl}$ is the set of possible states of that data memory, and
$\SText{\aw,\wl}$ is the set of possible transformations on those
states.

Let $\aw,\ous \geq 0$ and $\wl,\iss,\ssb > 0$, and let $\waf \in \Bool$
be such that $\waf = \False$ if $\aw = 0$.
Then the \emph{thread powered function class} with parameters $\aw$,
$\wl$, $\ous$, $\iss$, $\ssb$ and $\waf$, written
$\TPFC(\aw,\wl,\ous,\iss,\ssb,\waf)$, is the subset of $\SText{\aw,\wl}$
that is defined as follows:
\begin{ldispl}
T \in \TPFC(\aw,\wl,\ous,\iss,\ssb,\waf)
\\ {}
\Iff
\Exists{\Adm \subseteq \BAct}
 {{} \\ \quad \phantom{{} \Iff}
  \Exists{H \in \MISA(\aw,\wl,\ous,1,1,\Adm)}
   {{} \\ \qquad \phantom{{} \Iff}
    \Exists{p \in \Tfin(A_H)}
     {{} \\ \qquad\quad \phantom{{} \Iff}
      \bigl(\card(\Adm) = \iss \And \card(\Res(p)) \leq \ssb \And
      {} \\ \qquad\quad \phantom{{} \Iff} \phantom{\bigl(}
       \Forall{S \in \cS_H}
        {{} \\ \qquad\qquad \phantom{{} \Iff} \phantom{\bigl(}
         \bigl(\bigl(\waf = \False \Implies
           T(S \restr \Mext{\aw,\wl}) =
           (\apply{p}{H}{S}) \restr \Mext{\aw,\wl}\bigr) \And
         {} \\ \qquad\qquad \phantom{{} \Iff} \phantom{\bigl(\bigl(}
          \bigl(\waf = \True \Implies
         {} \\ \qquad\qquad \phantom{{} \Iff}
               \phantom{\bigl(\bigl(\bigl(}\;\;
           T(S \restr \Mext{\aw,\wl}) \restr \Mext{\aw-1,\wl} =
           (\apply{p}{H}{S}) \restr
           \Mext{\aw-1,\wl}\bigr)\bigr)}\bigr)}}}\;.
\end{ldispl}%
We say that $\TPFC(\aw,\wl,\ous,\iss,\ssb,\waf)$ is \emph{complete} if
$\TPFC(\aw,\wl,\ous,\linebreak[2]\iss,\ssb,\waf) = \SText{\aw,\wl}$.

The following theorem states that $\TPFC(\aw,\wl,\ous,\iss,\ssb,\waf)$
is complete if $\ous = 2^\aw \cdot \wl + \aw + 1$, $\iss = 5$ and
$\ssb = 8$.
Because $2^\aw \cdot \wl$ is the data memory size, i.e.\ the number of
bits that the data memory contains, this means that completeness can be
obtained with $5$ data manipulation instructions and threads whose
number of states is less than or equal to $8$ by taking the operating
unit size slightly greater than the data memory size.
\begin{theorem}
\label{theorem-completeness}
Let $\aw \geq 0$, $\wl > 0$ and $\waf \in \Bool$, and let
$\dms = 2^\aw \cdot \wl$.
Then $\TPFC(\aw,\wl,\dms + \aw + 1,5,8,\waf)$ is complete.
\end{theorem}
The idea behind the proof of Theorem~\ref{theorem-completeness} given
below is that first the content of the whole data memory is copied data
memory element by data memory element via the load data register to the
operating unit, after that the intended state transformation is applied
to the copy in the operating unit, and finally the result is copied back
data memory element by data memory element via the store data register
to the data memory.
The data manipulation instructions used to accomplish this are an
initialization instruction, a pre-load instruction, a post-load
instruction, a pre-store instruction, and a transformation instruction.
The pre-load instruction is used to update the load address register
before a data memory element is loaded, the post-load instruction is
used to store the content of the load data register to the operating unit
after a data memory element has been loaded, and the pre-store
instruction is used to update the store address register and to load the
content of the store data register from the operating unit before a data
memory element is stored.
The transformation instruction is used to apply the intended state
transformation to the copy in the operating unit.
\begin{proof}[Proof of Theorem~\ref{theorem-completeness}]
For convenience, we define
\begin{ldispl}
\begin{aeqns}
\MMData & = & \set{\mdata{\wl}(i) \where i \in [0,2^\aw-1]}\;,
\\
\MMOU \phantom{\langle i \rangle}
        & = & \set{\mou(j) \where j \in [0,\dms+\aw]}\;,
\\
\MMOUd & = & \set{\mou(j) \where j \in [0,\dms-1]}\;,
\\
\MMOUa & = & \set{\mou(j) \where j \in [\dms,dms + \aw]}\;,
\\
\MMOUdw{i} & = &
\set{\mou(j) \where j \in [i \cdot \wl,(i+1) \cdot \wl-1]}\;,
\mathrm{\;for\;} i \in [0,2^{\aw}-1]\;,
\\
\ldr & = & \mld{\wl}(0)\;,
\\
\sdr & = & \msd{\wl}(0)\;,
\\
\lar & = & \mla{\aw}(0)\;,
\\
\sar & = & \msa{\aw}(0)\;.
\end{aeqns}
\end{ldispl}%
We have that $\MMOUd = \Union_{i \in [0,2^{\aw}-1]} \MMOUdw{i}$ and
$\MMOU = \MMOUd \union \MMOUa$.

We have to deal with the binary representations of natural numbers in
the operating unit.
The set of possible binary representations of natural numbers in the
operating unit is
\begin{ldispl}
\cR = \Union_{n \in [0,\dms+\aw],m \in [n,\dms+\aw]}
       \set{R \where
            \funct{R}{\set{\mou(i) \where i \in [n,m]}}{\set{0,1}}}\;.
\end{ldispl}%
For each $R \in \cR$, the natural number of which $R$ is a binary
representation is given by the function $\funct{\bn}{\cR}{\Nat}$ that is
defined as follows:
\begin{ldispl}
\bn(R) =
\sum_{i \mathrm{\;s.t.\;} \mou(i) \in \dom(R)}
 R(\mou(i)) \cdot 2^{i - \min\set{j \where \mou(j) \in \dom(R)}}\;.
\end{ldispl}%

To prove the theorem, we take a fixed but arbitrary
$T \in \SText{\aw,\wl}$ and show that
$T \in \TPFC(\aw,\wl,\dms + \aw + 1,5,8,\waf)$.

We take $\Adm = \set{\init,\prel,\postl,\pres,\trf}$, and we take
$H = \tup{M,B,\cS,\cO,A,\Eval{\ph}} \in
 \MISA(\aw,\wl,\dms + \aw + 1,1,1,\Adm)$
such that $O_\init$, $O_\prel$, $O_\postl$, $O_\pres$, and $O_\trf$ are
the unique functions from $\cS$ to $\cS$ such that for all $S \in \cS$:
\begin{small}
\begin{ldispl}
\begin{aeqns}
O_\init(S) \restr (M \diff (\MMOUa \union \set{\replr})) & = &
S \restr (M \diff (\MMOUa \union \set{\replr}))\;,
\\
\bn(O_\init(S) \restr \MMOUa) & = & 0\;,
\\
O_\init(S)(\replr) & = & \True\;,
\end{aeqns}
\end{ldispl}%
\begin{ldispl}
\begin{aeqns}
\multicolumn{4}{@{}l@{}}
 {O_\prel(S) \restr
  (M \diff (\MMOUa \union \set{\lar} \union \set{\replr}))}
\\ & = &
\multicolumn{2}{@{}l@{}}
 {S \restr
  (M \diff (\MMOUa \union \set{\lar} \union \set{\replr}))\;,}
\\
\bn(O_\prel(S) \restr \MMOUa) & = & \bn(S \restr \MMOUa) + 1\;
 & \mif \bn(S \restr \MMOUa) < 2^\aw\;,
\\
O_\prel(S) \restr \MMOUa & = & S \restr \MMOUa
 & \mif \bn(S \restr \MMOUa) \geq 2^\aw\;,
\\
O_\prel(S)(\lar) & = & \bn(S \restr \MMOUa)
 & \mif \bn(S \restr \MMOUa) < 2^\aw\;,
\\
O_\prel(S)(\lar) & = & S(\lar)
 & \mif \bn(S \restr \MMOUa) \geq 2^\aw\;,
\\
O_\prel(S)(\replr) & = & \True
 & \mif \bn(S \restr \MMOUa) < 2^\aw\;,
\\
O_\prel(S)(\replr) & = & \False
 & \mif \bn(S \restr \MMOUa) \geq 2^\aw\;,
\eqnsep
\multicolumn{4}{@{}l@{}}
 {O_\postl(S) \restr
  (M \diff (\MMOUdw{\bn(S \restr \MMOUa)} \union \set{\replr}))}
\\ & = &
\multicolumn{2}{@{}l@{}}
 {S \restr
  (M \diff (\MMOUdw{\bn(S \restr \MMOUa)} \union \set{\replr}))\;,}
\\
\bn(O_\postl(S) \restr \MMOUdw{\bn(S \restr \MMOUa)}) & = & S(\ldr)\;,
\\
O_\postl(S)(\replr) & = & \True\;,
\eqnsep
\multicolumn{4}{@{}l@{}}
 {O_\pres(S) \restr
  (M \diff (\MMOUa \union \set{\sar,\sdr} \union \set{\replr}))}
\\ & = &
\multicolumn{2}{@{}l@{}}
 {S \restr
  (M \diff (\MMOUa \union \set{\sar,\sdr} \union \set{\replr}))\;,}
\\
\bn(O_\pres(S) \restr \MMOUa) & = & \bn(S \restr \MMOUa) + 1\;
 & \mif \bn(S \restr \MMOUa) < 2^\aw\;,
\\
O_\pres(S) \restr \MMOUa & = & S \restr \MMOUa
 & \mif \bn(S \restr \MMOUa) \geq 2^\aw\;,
\\
O_\pres(S)(\sar) & = & \bn(S \restr \MMOUa)
 & \mif \bn(S \restr \MMOUa) < 2^\aw\;,
\\
O_\pres(S)(\sar) & = & S(\sar)
 & \mif \bn(S \restr \MMOUa) \geq 2^\aw\;,
\\
O_\pres(S)(\sdr) & = & \bn(S \restr \MMOUdw{\bn(S \restr \MMOUa)})\;
 & \mif \bn(S \restr \MMOUa) < 2^\aw\;,
\\
O_\pres(S)(\sdr) & = & S(\sdr)
 & \mif \bn(S \restr \MMOUa) \geq 2^\aw\;,
\\
O_\pres(S)(\replr) & = & \True
 & \mif \bn(S \restr \MMOUa) < 2^\aw\;,
\\
O_\pres(S)(\replr) & = & \False
 & \mif \bn(S \restr \MMOUa) \geq 2^\aw\;,
\eqnsep
O_\trf(S) \restr (M \diff (\MMOU \union \set{\replr})) & = &
S \restr (M \diff (\MMOU \union \set{\replr}))\;,
\\
O_\trf(S) \restr \MMOUd & = & T'(S \restr \MMOUd)\;,
\\
\bn(O_\trf(S) \restr \MMOUa) & = & 0\;,
\\
O_\trf(S)(\replr) & = & \True\;,
\end{aeqns}
\end{ldispl}%
\end{small}
where $T'$ is the unique function from
$\set{S \restr \MMOUd \where S \in \cS}$ to
$\set{S \restr \MMOUd \where S \in \cS}$ such that, for all
$S_\nm{ou}^d \in \set{S \restr \MMOUd \where S \in \cS}$, there exists an
$S_\nm{data} \in \set{S \restr \MMData \where S \in \cS}$ such that:
\begin{ldispl}
\Forall{i \in [0,2^\aw-1]}{{}}
\\ \quad
 (\bn(S_\nm{ou}^d \restr \MMOUdw{i}) = S_\nm{data}(\MMData[i]) \And {}
\\ \quad \phantom{(}
  \bn(T'(S_\nm{ou}^d) \restr \MMOUdw{i}) = T(S_\nm{data})(\MMData[i]))\;.
\end{ldispl}%

Moreover, we take $p \in \Tfin(A)$ such that $p = \rec{X}{E}$ with $E$
consisting of the following equations:
\pagebreak[2]
\begin{ldispl}
\begin{aeqns}
X & = & \init \bapf Y\;,
\\
Y & = & \pcc{(\load{:}0 \bapf \postl \bapf Y)}{\prel}{(\trf \bapf Z)}\;,
\\
Z & = & \pcc{(\store{:}0 \bapf Z)}{\pres}{\Stop}\;.
\end{aeqns}
\end{ldispl}%

Let $S \in \cS$ and
let $\tup{p_i,S_i}$ be the $(i+1)$st element in the full path of
$\tup{\rec{X}{E},S}$ on $H$ whose first component equals
$\rec{X}{E}$, $\rec{Y}{E}$, $\rec{Z}{E}$ or $\Stop$.
Then it is easy to prove by induction on $i$ that
\begin{ldispl}
i = 1 \Implies
p_i = \rec{Y}{E} \And
\bn(S_i \restr \MMOUa) = 0\;,
\\
i \in [2,2^\aw + 2] \Implies {}
\\ \quad
p_i = \rec{Y}{E} \And {}
\\ \quad
\Forall{j \in [0,i - 2]}{\bn(S_i \restr \MMOUdw{j}) = S(\MMData[j])}
 \And {}
\\ \quad
\bn(S_i \restr \MMOUa) = i - 1\;,
\\
i = 2^\aw + 3 \Implies {}
\\ \quad
p_i = \rec{Z}{E} \And {}
\\ \quad
\Forall{j \in [0,2^\aw - 1]}
 {\bn(S_i \restr \MMOUdw{j}) = T(S \restr \MMData)(\MMData[j])}
  \And {}
\\ \quad
\bn(S_i \restr \MMOUa) = 0\;,
\\
i \in [2^\aw + 4,2^{\aw+1} + 4] \Implies {}
\\ \quad
p_i = \rec{Z}{E} \And {}
\\ \quad
\Forall{j \in [0,i - (2^\aw + 4)]}
 {S_i(\MMData[j]) = T(S \restr \MMData)(\MMData[j])} \And {}
\\ \quad
\bn(S_i \restr \MMOUa) = i - (2^\aw + 3) \;,
\\
i = 2^{\aw+1} + 5 \Implies
p_i = \Stop \And
S_i \restr \MMData = T(S \restr \MMData)\;.
\end{ldispl}%
Hence,
$(\apply{\rec{X}{E}}{H}{S}) \restr \MMData =
 T(S \restr \MMData)$.
That is, $T$ can be achieved by applying $\rec{X}{E}$ to $H$.
\end{proof}

As a corollary of the proof of Theorem~\ref{theorem-completeness}, we
have that in the case where $\waf = \True$ completeness can also be
obtained if we take about half the external memory size as the operating
unit size.
\begin{corollary}
\label{corollary-completeness-waf}
Let $\aw > 0$ and $\wl > 0$, and let $\ems = 2^{\aw-1} \cdot \wl$.
Then $\TPFC(\aw,\wl,\ems + \aw,5,8,\True)$ is complete.
\end{corollary}

As a corollary of the proofs of Theorems~\ref{theorem-reducing-ous}
and~\ref{theorem-completeness}, we have that completeness can even be
obtained if we take zero as the operating unit size. However, this may
require quite a large number of data manipulation instructions and
threads with quite a large number of states.
\begin{corollary}
\label{corollary-completeness-small-ous}
Let $\aw \geq 0$ and $\wl > 0$, let $\waf \in \Bool$ be such that
$\waf = \False$ if $\aw = 0$, and let $\dms = 2^\aw \cdot \wl$.
Then
$\TPFC(\aw,\wl,0,5 \cdot 4^{\dms + \aw + 1},
                 8 \cdot 6^{\dms + \aw + 1},\waf)$
is complete.
\end{corollary}

\section{On Incomplete Thread Powered Function Classes}
\label{sect-TPF-incomplete}

From Corollary~\ref{corollary-completeness-small-ous}, we know that it
is possible to achieve all transformations on the states of the external
memory of a strict load/store Maurer ISA with given address width and
word length even if the operating unit size is zero.
However, this may require quite a large number of data manipulation
instructions and threads with quite a large number of states.
This raises the question whether the operating unit size of the ISAs,
the size of the instructions set of the ISAs and the maximal number of
states of the threads can be taken such that it is impossible to achieve
all transformations on the states of the external memory.

Below, we will give a theorem concerning this question, but first we
give a lemma that will be used in the proof of that theorem.
\begin{lemma}
\label{lemma-incompleteness}
Let $\aw > 1$ and $\wl,\ous,\iss,\ssb > 0$, and
let $\ems = 2^{\aw-1} \cdot \wl$.
Then $\TPFC(\aw,\wl,\ous,\iss,\ssb,\True)$ is not complete if
$\ous \leq \ems/2$ and $\iss \leq 2^{\ems/2}$ and there are no more than
$2^\ems$ threads that can be applied to the members of
$\Union_{\Adm \subseteq \BAct} \MISA(\aw,\wl,\ous,1,1,\Adm)$.
\end{lemma}
\begin{proof}
We have that, if the operating unit size is no more than $\ems/2$, no
more than
\begin{ldispl}
{(2^{\ems/2})}^{(2^{\ems/2})}
\end{ldispl}%
transformations on the states of the operating unit can be associated
with one data manipulation instruction.
Because the load instruction and the store instruction are interpreted
in a fixed way, it follows that, if there are no more than $2^{\ems/2}$
data manipulation instructions, no more than
\begin{ldispl}
{\Bigl({(2^{\ems/2})}^{(2^{\ems/2})}\Bigr)}^{(2^{\ems/2})}
\end{ldispl}%
transformations on the states of the external memory can be achieved
with one thread.
Hence, if no more than $2^\ems$ threads can be applied, no more than
\begin{ldispl}
{\Bigl({(2^{\ems/2})}^{(2^{\ems/2})}\Bigr)}^{(2^{\ems/2})}
\cdot 2^\ems
\end{ldispl}%
transformations on the states of the external memory can be achieved.
Using elementary arithmetic, we easily establish that
\begin{ldispl}
{\Bigl({(2^{\ems/2})}^{(2^{\ems/2})}\Bigr)}^{(2^{\ems/2})}
\cdot 2^\ems <
{(2^\ems)}^{(2^\ems)}\;.
\end{ldispl}%
It follows that $\TPFC(\aw,\wl,\ous,\iss,\ssb,\True)$ is not complete
because ${(2^\ems)}^{(2^\ems)}$ is the number of transformations that
are possible on the states of the external memory.
\end{proof}
In Lemma~\ref{lemma-incompleteness}, the bound on the number of threads
that can be applied seems to appear out of the blue.
However, it is the number of threads that can at most be represented in
the internal memory: with the most efficient representations we cannot
have more than one thread per state of the internal memory.

The following theorem states that $\TPFC(\aw,\wl,\ous,\iss,\ssb,\True)$
is not complete if the operating unit size is not greater than half the
external memory size, the instruction set size is not greater than
$2^\wl - 4$, and the maximal number of states of the threads is not
greater than $2^{\aw-2}$.
Notice that $2^\wl$ is the number of instructions that can be
represented in memory elements with word length $\wl$ and that
$2^{\aw-2}$ is half the number of memory elements in the internal
memory.
\begin{theorem}
\label{theorem-incompleteness}
Let $\aw,\wl > 1$ and $\ous,\iss,\ssb > 0$, and
let $\ems = 2^{\aw-1} \cdot \wl$.
Then $\TPFC(\aw,\wl,\ous,\iss,\ssb,\True)$ is not complete if
$\ous \leq \ems/2$ and $\iss \leq 2^\wl - 4$ and $\ssb \leq 2^{\aw-2}$.
\end{theorem}
\begin{proof}
The number of threads with at most $\ssb$ states does not exceed
\begin{ldispl}
\bigl((\iss + 2) \cdot \ssb^2 + 2\bigr)^\ssb
\end{ldispl}%
(recall that there is also one load instruction and one store
instruction).
Because $\ssb > 0$,
\begin{ldispl}
\bigl((\iss + 2) \cdot \ssb^2 + 2\bigr)^\ssb\ \leq
\bigl((\iss + 4) \cdot \ssb^2\bigr)^\ssb\;.
\end{ldispl}%
Using elementary arithmetic, we easily establish that
\begin{ldispl}
\bigl((\iss + 4) \cdot \ssb^2\bigr)^\ssb \leq 2^\ems\;.
\end{ldispl}%
Consequently, there are no more than $2^\ems$ threads with at most
$\ssb$ states.
From this, and the facts that $\ous \leq \ems/2$ and
$\iss < 2^{\ems/2}$, it follows by Lemma~\ref{lemma-incompleteness} that
$\TPFC(\aw,\wl,\ous,\iss,\ssb,\True)$ is not complete.
\end{proof}

\section{Conclusions}
\label{sect-conclusions}

In~\cite{BM05f,BM06b}, we have worked at a formal approach to
micro-architecture design based on Maurer machines and basic thread
algebra.
In those papers, we made hardly any assumption about the instruction set
architectures for which new micro-architectures are designed, but we put
forward strict load/store Maurer instruction set architectures as
preferable instruction set architectures.
In the current paper, we have established general properties of strict
load/store Maurer instruction set architectures.

Some of these properties presumably belong to the non-trivial insights
of practitioners involved in the design of instruction set
architectures:
\begin{iteml}
\item
Theorem~\ref{theorem-reducing-ous} clarifies the ever increasing
operating unit size.
In principle, it is possible to undo an increase of the operating unit
size originating from matters such as more advanced pipelined
instruction processing, more advanced superscalar instruction issue, and
more accurate floating point calculations.
However, the price of that is prohibitive measured by the increase of
the size of the instruction set needed to produce the same state changes
on the data memory concerned, the number of steps needed for those state
changes, etc.
\item
Theorem~\ref{theorem-completeness} explains the existence of programming
to a certain extent.
In principle, it is possible to achieve each possible transformation on
the states of the data memory of an instruction set architecture with
a very short simple program.
However, that requires instructions dedicated to the different
transformations and an operating unit size that is broadly the same as
the data memory size.
\item
Theorem~\ref{theorem-incompleteness} points out the limitations imposed
by existing instruction set architectures.
Even if we take the size of the operating unit, the size of the
instruction set, and the maximal number of states of the threads applied
much larger than found in practice, not all possible transformations on
the states of the data memory concerned can be achieved.
\end{iteml}
If we remove the conditions imposed on the data manipulation
instructions of strict load/store Maurer instruction set architectures,
then Theorems~\ref{theorem-reducing-ous}, \ref{theorem-completeness},
and~\ref{theorem-incompleteness} still go through.
These conditions originate from~\cite{BM06b}, where they were introduced
to allow for exploiting parallelism, with the purpose to speed up
instruction processing, to the utmost.

In this paper, we have taken the view that transformations on the states
of the external memory are achieved by applying threads.
We could have taken the less abstract view that transformations on the
states of the external memory are achieved by running stored programs.
This would have led to complications which appear to be needless because
only the threads that are represented by those programs are relevant to
the transformations on the states of the external memory that are
achieved.
However, in the case of Theorem~\ref{theorem-incompleteness}, the bounds
that are imposed on strict load/store Maurer instruction set
architectures and threads induce an upper bound on the number of threads
that can be applied.
This raises the question whether the number of threads that can be
represented in the internal memory of the instruction set architectures
concerned is possibly higher than the upper bound on the number of
threads that can be applied.
This question can be answered in the negative: the upper bound is the
number of threads that can at most be represented in the internal
memory.
The upper bound is in fact much higher than the number of threads that
can in practice be represented in the internal memory (cf.\ the remark
after Lemma~\ref{lemma-incompleteness}).
This means that Theorem~\ref{theorem-incompleteness} demonstrates that
load/store instruction set architectures impose restrictions on the
expressiveness of computers, also if we take the view that
transformations on the states of the external memory are achieved by
running stored programs.

One of the options for future work is to improve upon the results given
in this paper.
For example, we know from Theorem~\ref{theorem-completeness} that, in
order to obtain completeness with $5$ data manipulation instructions and
threads whose number of states is less than or equal to $8$, it is
sufficient to take the operating unit size slightly greater than the
data memory size.
However, we do not yet know what the smallest operating unit size is
that will do.
Another option for future work is to establish results that bear upon
the use of half the data memory as internal memory.
No such results are given in this paper.
In Lemma~\ref{lemma-incompleteness}, it is assumed that half the data
memory is used as internal memory because it provides a good case for
the number taken as the maximal number of threads that can be applied.
However, Lemma~\ref{lemma-incompleteness}, as well as
Theorem~\ref{theorem-incompleteness}, goes through without internal
memory.

The speed with which transformations on the states of the external
memory are achieved depends largely upon the way in which the strict
load/store Maurer instruction set architecture in question is
implemented.
This hampers establishing general results about it.
However, the speed with which transformations on the states of the
external memory are achieved depends also on the volume of data transfer
needed between the external memory and the operating unit.
Establishing a connection between this volume and the parameters of
thread powered function classes is still another option for
future work.

\bibliographystyle{spmpsci}
\bibliography{CA}

\end{document}